# Understanding Knowledge Leakage & BYOD (Bring Your Own Device): A Mobile Worker Perspective

**Carlos A. Agudelo**
School of Engineering
University of Melbourne
Melbourne, Australia
Email: cagudelo@student.unimelb.edu.au

**Rachelle Bosua**
School of Engineering
University of Melbourne
Melbourne, Australia
Email: rachelle.bosua@unimelb.edu.au

**Atif Ahmad**
School of Engineering
University of Melbourne
Melbourne, Australia
Email: atif@unimelb.edu.au

**Sean B. Maynard**
School of Engineering
University of Melbourne
Melbourne, Australia
Email: seanbm@unimelb.edu.au

## Abstract

Knowledge sharing drives innovation and the opportunity to develop a sustainable competitive advantage. However, in the extant knowledge management and information security literature, leakage from sharing activities is neglected. The risk of knowledge leakage is exacerbated with the pervasive use of mobile devices and the adoption of BYOD (Bring Your Own Device). Thus, this research-in-progress paper examines the role of the behaviour of mobile workers that engage in accidental knowledge leakage through the use of BYOD. We use the Decomposed Theory of Planned Behaviour (DTPB) to explain the causes behind this phenomenon and how it negatively impacts organization's competitive advantage. The contributions of this study are the following. First, it posits that the reasons of knowledge leakage by mobile workers through BYOD can be explained using DTPB. Second, the paper proposes a conceptual framework for research based on DTPB constructs whilst adding other variables such as BYOD and mobile device usage context. Finally, the conceptual study outlines the potential contributions and implications of this research.

## 1   Introduction

Knowledge is a key strategic organizational asset that must be protected since it is a competitive resource and source of innovation (Ahmad, Bosua, et al. 2014). However, despite its relevance to organizations, knowledge protection has been neglected due to the fact that it is considered a barrier to knowledge sharing and creation workflows within organizations; moreover neglecting knowledge protection can cause the replication of ideas by external organizations hindering the exploitation of innovation (Shedden et al. 2011). The challenge of finding a balance between knowledge sharing and protection is further amplified by mobile technologies and social media that facilitate knowledge sharing and increase the risk of knowledge leakage(Manhart and Thalmann 2015).

Knowledge leakage, which is part of the risks addressed by knowledge protection, is defined in this paper as the ***accidental or deliberate loss or unauthorized transfer of organizational knowledge intended to stay within a firm's boundary that may weaken the*** ***competitiveness and industrial position of the organization*** (Annansingh 2006; Frishammar et al. 2015).

Knowledge leakage can transpire from the disclosure of sensitive information as the receiver or competitor (external knowledge bearer) of this information may infer knowledge from what it is being leaked. This situation is exacerbated by the use of mobile devices (BYOD) by mobile workers, where the leaked knowledge has been previously codified (explicit) into artefacts, such as documents, pictures, videos and audio recordings. The pervasive use of BYOD has become a conduit for information exchange not only for personal purposes but also for professional reasons, exposing the organization to the risk of knowledge leakage. This situation is even more challenging for ICT departments to control, as it extends the boundaries of the organization's network and immediate control of organization's security staff (Ahmad et al. 2014). Consequently, this risk can have negative effects on an organization's competitive advantage, reputation and financial situation (Abdul Molok et al. 2010).

By the same token, employees pose security risks due to their legitimate access to facilities and information, knowledge of the organization and the location of valuable assets; this is known as the insider threat and it constitutes an even greater risk with the use of online social network and the technical capabilities of mobile devices (Colwill 2009). Whilst employee's behaviour can be categorized in to accidental and malicious, previous research (Colwill 2009) has shown that accidental behaviour exerted by insiders may have greater potential for harm to organizations than malicious insider attacks, and it is also more frequent. Therefore, organizations should realize that their people may be their greatest vulnerability and technical controls are not enough to prevent this behaviour (Ahmad, Maynard, et al. 2014). This requires a focus on human factors, perception and expectations. Consequently, this research will concentrate on accidental behaviour that leads to knowledge leakage by employees.

While confidentiality, integrity and availability are important aspects to knowledge security in knowledge management literature, this study will only focus on the *confidentiality* perspective as it is more relevant to knowledge leakage than the other two aspects. Therefore, it is important to investigate how organizations address this phenomenon inside and outside their boundaries, posing the following questions:

1. *Why do mobile workers unintentionally leak organizational knowledge through BYOD in various usage contexts?*
2. *How can organizations protect their knowledge from being leaked by mobile workers through BYOD?*

To answer these questions, this conceptual paper proposes using DTPB (Decomposed Theory of Planned Behaviour) as a theoretical framework to explain unintentional behaviour, adding other variables such as BYOD and mobile device usage context. Understanding the reasons behind this behaviour can assist organizations in developing more effective policies and SETA (Security Education Training and Awareness) programs to address this issue.

The rest of this paper is structured as follows: First, it provides the salient concepts found in the key background literature. Second, the study presents the conceptual framework proposed in this research. Finally, the study outlines the potential contributions and implications of this study.

## 2   Key Background Literature

This section reviews the relevant literature areas. It first examines work in knowledge along with knowledge leakage. It then discusses the concepts of mobility, ownership and context before presenting work on mobile worker behaviour. Finally, it introduces the Decomposed Theory of Planned Behaviour.

### 2.1   Knowledge

This research-in-progress paper adopts the definition of knowledge given by Davenport and Prusak (1998):

*"Knowledge is a fluid mix of framed experience, values, contextual information, and expert insight that provides a framework for evaluating and incorporating new experiences and information. It originates and is applied in the minds of knowers. In organizations, it often becomes embedded not only in documents or repositories but also in organizational routines, processes, practices, and norms."*

According to this definition, knowledge is complex as it is a mixture of various elements, it is intuitive and, therefore, hard to capture. Moreover, it is within people, and as such, it may be unpredictable and intangible. Knowledge derives from information and to turn it into knowledge, people must do all the work (Davenport and Prusak 1998). Although knowledge is further divided into tacit (present in employee's minds) and explicit (knowledge that has been codified into artefacts) (Nonaka 1991), this study will only focus on explicit knowledge leakage, since its disclosure is more likely to occur than tacit knowledge leakage, such as when key personnel leave the organization to a competitor (Frishammar et al. 2015).

While the world is shifting into the digital era, information and knowledge are key strategic assets (Bollinger and Smith 2001) for an organisation to achieve sustainable competitive advantage, innovation and value creation (Nonaka and Toyama 2003; Sveiby 1997). Likewise, MacDougall & Hurst (2005) contend that the adoption of knowledge workers, employees who produce value by utilising their knowledge rather than physical labour, allows organisations to develop their knowledge assets. These individuals perform work based on their information assets for the coordination and management of organisational activities (Sorensen et al. 2008). Ristovska et al. (2012) also focus on the importance of knowledge embedded in the knowledge worker as it is an organisational asset for achieving sustainable competitive advantage which can be materialized into documentation and organizational processes. Knowledge in this sense is the information residing in the mind of the knowledge worker, personalised by the individual based on facts, procedures, concepts, interpretations, ideas, observations and judgements which is codified into artefacts such as documentation, processes and guidelines (Alavi and Leidner 2001).

## 2.2 Knowledge Leakage Risk

Although knowledge management can help an organisation build and maintain their competitive advantage, knowledge leakage on the other hand may result in the loss of the organisation's competitiveness. Annansingh (2006) defines knowledge leakage as "the possibility of information or knowledge that is critical to the organisation being lost or leaked – whether deliberately or unintentionally to a competitor or unauthorised personnel". Although knowledge being lost due to a lack of knowledge management procedures is also defined as knowledge leakage (Nunes et al. 2006), the focus in this study will be on knowledge leakage directly or indirectly caused by knowledge workers when performing knowledge work, as it is the most challenging channel of leakage for organisations to control (Nunes et al. 2006).

The protection of organizational knowledge is fundamental to maintain competitive advantage in knowledge intensive organizations; furthermore, the importance of expertise in organisations relies heavily on the exercising of specialist knowledge and competencies, or alternatively, the management of organizational competencies and capabilities which belong to employees or knowledge workers (Ahmad, Bosua, et al. 2014; Blackler 1995).

## 2.3 Mobility

The risk of knowledge leakage is significantly elevated as the technology environment evolves. Changes in the technology environment have propelled transformation of how employees perform knowledge work, thus, influencing and facilitating the way users behave and create knowledge. As Green (2002) claims, the change in the technology environment is enabled through the development of the modern metropolis embedded with strong telecommunication infrastructures. An increase of strength and reliability of Wi-Fi and cellular networks, as well as the increased availability of unsecured public Wi-Fi hotspots, allows users to stay connected in many different environments and situations between the home and workplace. The shift from hard-disk storage to cloud storage is a significant example of change in technology infrastructure as well as the adoption and appropriation of boundary-spanning technologies (Ahmad, Bosua, et al. 2014). These infrastructure changes enable technologies such as mobile devices to move and operate seamlessly through networks for both work and personal use (Sorensen et al., 2008).

Globalisation and economies of scale have stimulated the phenomenon of consumerization which describes the wider adoption of technologies such as mobile devices due to lower costs of production and distribution (Moschella et al. 2004). Mobile devices provide users with technological capabilities to perform work outside the organisation through access to corporate assets such as emails, enabling higher productivity which has driven the adoption of BYOD (Bring Your Own Device) policies within organizations (Ghosh and Rai 2013; Sarker and Wells 2003). Further, supported by changes in technological infrastructure, mobile devices are becoming highly capable and personally located devices near users which are always on and are connected to a variety of networks extending the corporate environment.

## 2.4 Ownership

As adoption rates of mobile technologies proliferate, employees are increasingly purchasing their own personal mobile devices and incorporating them into their organizations. From this emerging trend, ownership provides the concepts of Employee Managed Personal Devices (EMPD), where mobile devices are carried by the employee who has a high degree of control to modify and use their devices as they see fit. Corporate managed personal devices (CMPD), on the other hand are devices controlled by the organisation, although still carried and used personally by the employee inside and outside of the workplace (Sindhu et al. 2010).

The concept of EMPD and CMPD introduces a trade-off of ownership between organisations and individuals for the control and usage of mobile devices. On one hand, personally owned devices can be used in a corporate environment for corporate work where there is no control of the device from the organisation. On the other hand, corporate devices can be used by employees, within the restrictions of corporate policy. The use of mobile devices, whether corporate or employee owned, extends the infrastructure perimeter of an organisation to where the user travels during work as well as in their private time (Sindhu et al. 2010). This extension greatly increases exposure of the device to security risks of knowledge leakage. The focus of this research, however, is on situations of knowledge leakage where employees are in control of the device (e.g. using the device on the train, park or the airport to perform work), as this is more challenging for organizations to control and brings the user behaviour dimension to the problem.

Since our definition of knowledge leakage focuses on leakage as a result of the employee's usage of the device for mobile knowledge work, situations in which employees have little control of (such as a third-party hacking attempts) will be out of scope as they may not necessarily be caused by the user's interaction with the device. Misuse of the device such as clicking on suspicious email links (i.e. phishing) however, is considered within the scope of knowledge leakage as it is the misuse of the device in control of the user. Knowledge Leakage can occur in the physical environment (e.g. Shoulder surfing) as well as technical environment (e.g. Man in the Middle attacks as a result of users connecting to unencrypted public Wi-Fi access points).

## 2.5 Context

Although knowledge leakage is enabled by the employee in control of the mobile device, there are multiple environmental factors that affect the use of mobile devices for knowledge work. Nonaka & Toyama (2003) suggest that knowledge creation, sharing and distribution are achieved through the interactions between the individual, the organisation and the environment. The environment influences the individual while, at the same time, the individuals are continuously recreating their environment through their social interactions. This proposes that social factors in human interactions constantly change the environment in which knowledge is created. Nonaka & Toyama (2003) developed a model of knowledge creation in order to explain the conversion of knowledge through interactions between individuals, groups of individuals, organisations and the environment. This model not only highlights the importance of the environmental and organisational circumstances around an individual, it also highlights the importance of the social environment where individuals interact within groups to obtain information.

These environments are referred to in the literature, from a mobile device perspective, as the "context" of the mobile device usage. Table 1 summarises the context taxonomy across the literature. Understanding the context of the use of these mobile devices in these different settings (environmental, organisational and social) is important to assess the overall security risk of the device as the potential enabler or medium in which knowledge leakage can occur in conjunction with the user and the environment within which the device is used.

| Context | Author | Description |
| --- | --- | --- |
| Environmental Context | Kofod-Petersen & Cassens (2006); Nieto, Botía, & Gómez-Skarmeta, (2006) | The environmental context is defined as the conjunction of the following contexts: temporal context, spatial context, social context, technological context, and business context |

| Context | Author | Description |
|---|---|---|
| Personal Context | Kofod-Petersen & Cassens (2006) | The personal context provides the attributes of cognitive skills and draws on psychological and physiological contexts: psychological, goal, cognition, physiological, identity, actions |
| Social Context | Nieto et al., (2006) | Provides a social perspective of context, which captures the attributes of people (e.g. attitude, skills, and values) and the relationship of these people among each other and within the organisation structure. |
| Spatial Context | Kofod-Petersen & Cassens (2006) | Provides attributes of location and answers the question of where the interaction is conducted. The following are some constructs of this category: spatial objects, localization, location, season, weather, geography, routes, building |
| Temporal context | Abdoul Aziz Diallo (2012) | Temporal context is defined in terms of when the activity is performed: absolute date (year, month, day, hour, minutes, seconds), relative date (yesterday, tomorrow, next month, next year, etc.) |
| User Context | Abdoul Aziz Diallo (2012) | User context extends on personal context adding the technological dimension and the mobile device |
| Location Context | Abdoul Aziz Diallo (2012) | Location context is part of the spatial context and it is defined by: places, GPS location |
| Business Context | Abdoul Aziz Diallo (2012) | The business context supports the decision making process by assisting in the decision maker's situation awareness cognitive process, and taking in to consideration the following aspects: indicators, objectives, partners, competitors, market |
| Technological Context | Abdoul Aziz Diallo (2012) | Provides the technological and technical attributes such as: network connections, infrastructure, equipment, devices |
| *Organisational Context* | *Proposed Dimension in this study* | *Information Security Policies, SETA (Security Education Training and Awareness), Culture, Standards, organisational processes and procedures* |

*Table 1 Taxonomy of Contexts*

Mohamed et al. (2006) found that one of the key routes of knowledge leakage is people through social contexts of mobile usage. These routes include training courses, collaborations with universities, multi-disciplinary teams and temporary workers. Through social interactions in these different contexts, knowledge is shared or accessible to other users. Social context also includes the use of social networking platforms on the mobile device (Krishnamurthy and Wills 2010).

Thus, mobile device usage context will be referred to in this study as the combination of the following factors:
- The mobile knowledge worker working for the organization
- The device and its content
- The various surrounding dynamic environments within which the mobile device is used.

Due to the nature of mobile device usage, the context of a device transitions across many changes in technical, social and locational environments. Through the interactions of these dynamic contexts with one another, the risk of knowledge leakage also becomes dynamic. Thus, knowledge can be leaked through the technological and network context among others. As an illustration of this phenomenon, Astani et al (2013) found that a significant amount of employees from information sensitive industries such as banking, connected their mobile devices to unsecured public Wi-Fi networks which exposes the device to the security vulnerabilities of those networks and may be used as a medium for knowledge leakage. By simply changing the network connection to a public Wi-Fi network, these employees are drastically changing the environment and, therefore, their "mobile device usage context" in which the device is operating, changing the risk profile of their device.

Thus, it is the user, the environment, and the device along with its content combined that define the various contexts within which knowledge leakage risk changes (For instance, social context, technical context, personal context, etc.). These contexts are relevant to the usage of the device. If a user changes device, for example, then their overall context will change since the new one may not have the same functions as the previous one which can reduce the contexts that affect the device. Since the old device is no longer used by the user, various contexts such as social context and personal context no longer apply to it. This highlights the dynamic changes in knowledge leakage risks as the circumstances of how the knowledge worker uses their device change.

Additionally, people and objects are constantly moving in and out of contexts and the relevancy of these objects and people to the context are dynamic and hence the security threat of knowledge leaking is constantly changing. For example, if John is sitting in a coffee store reading his corporate emails from his tablet before heading into work and a new customer sits down behind John, John's risk context has changed as the customer may potentially read John's tablet screen (shoulder surfing). John then receives a phone call, which introduces a new person (caller) into the context, who then discusses the agenda of the morning meeting. This change in context now involves the surrounding people within earshot for potential knowledge leakage.

From the literature there have been many approaches to modelling the contextual information surrounding mobiles across many disciplines of Information Technology. Most of the research into the *contextual information* and *context* of mobile devices has been focused on the technical and computing issues (Benítez-Guerrero et al. 2012; Bradley, Nicholas A and Dunlop 2005; Kofod-Petersen and Cassens 2006; Schilit et al. 1994).

Likewise, Hofer et al. (2003) also extended and modelled these dimensions of context into device context (e.g. Device identifier and device type) and network context (e.g. network connection types) which were included as the technical context, in a more recent study, by Abdoul Aziz Diallo (2012). However, these studies failed to address the social context, neglecting the human perspective into the mobile contextual model, namely, user behaviour.

## 2.6 Mobile Worker Behaviour

By acknowledging the dynamic nature of knowledge leakage risk, it is important to understand the behaviour of the mobile knowledge worker to find the underlying reasons that explain why they perform work in risky contexts. Subsequently several behaviour theories have been explored through the lens of psychology, sociology and criminology. These theories have been summarized in Table 2.

| Theory | Reference | Summary |
|---|---|---|
| General Deterrence Theory (GDT) | (Straub 1990) | This criminology theory supports the use of information security policy, awareness training and preventive security systems as key deterrents to computer abuse. However, GDT does not explain the underlying factors as to why computer abuse occurs. |
| Theory of Reasoned Action (TRA) | (Ajzen et al. 1991) | Theory of Reasoned Action is a social psychology theory that mentions behaviour intention is the driver of behaviour and intentions are formed by a person's attitude and subjective norm or social influence. |
| Technology Acceptance Model (TAM) | (Davis, Bagozzi, & , 1989) | Technology Acceptance Model is an adaption of TRA. This theory specifies two beliefs, perceived usefulness and perceived ease of use, as the determinants of attitude that results to intentions and IT use |
| Theory of Planned Behaviour (TPB) | (Ajzen et al. 1991) | Theory of Planned behaviour is a psychology theory that expands on TRA with an additional construct called perceived behavioural control which means behaviour can happen even when people do not have a complete control over their behaviour. This supposes that people's behaviour can be accidental and intentional, and the behaviour is also caused by social influence and the person's attitude towards information system (IS) usage |

| Theory | Reference | Summary |
|---|---|---|
| Decomposed Theory of Planned Behaviour (DTPB) | (Taylor, S., & Todd 1995) | Decomposed Theory of Planned Behaviour combines TAM with TPB and decomposes the attitudinal, normative and control belief structures (see Figure 1).<br>While other theories mostly explain about behaviour that is intentional, DTPB posits that behaviour can be accidental and intentional since it is performed within and beyond the person's control.<br>In their study, it is contended that what influences the person to perform the behaviour of IT use is better explained by analysing the "multiple dimensions of subjective norms, and decomposing the perceived behavioural control dimension to evaluate self-efficacy along with technology and resource facilitating conditions" (Herath and Rao 2009, p. 108) |

*Table 2 Behavioural Theories*

## 2.7 Decomposed Theory of Planned Behaviour

After analysing these theories, this study found that DTPB is the most suitable to answer the research question: *Why do mobile workers unintentionally leak organizational knowledge through BYOD in various usage contexts?* Since DTPB explains that behaviour may occur intentionally and accidentally, in line with IS (Information Systems) findings that indicate that insider threats to organizations are mostly due to accidental behaviour (Colwill 2009; Taylor, S., & Todd 1995).

Furthermore, the perceived behavioural control construct exerts that behaviour can be accidental and intentional as it is performed within and beyond the person's control, determined by self-efficacy (perceived ability) and facilitating conditions. In other words, mobile workers using BYOD engage in behaviours that are intentional and accidental. For example, a mobile worker in an airport may inadvertently connect to a public Wi-Fi network and check confidential information concerning organization's trade secrets exposing this sensitive information (explicit codified knowledge) to outsiders, which may constitute a knowledge leakage incident to a competitor.

Understanding the reasons behind this behaviour will assist organizations in developing better policies and controls, therefore building on DTPB, both the intentional and accidental behaviour of mobile workers eliciting knowledge leakage risk through BYOD can be expressed as a function of these three factors: *attitude*, *subjective norms* or *social influence* and *perceived behavioural control*, along with their decomposed structures (Taylor and Todd 1995) as shown in Figure 1.

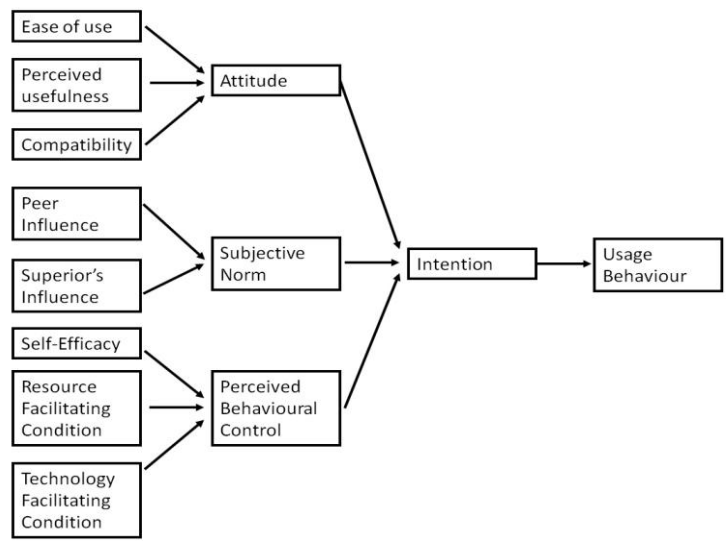

*Figure 1 Decomposed Theory of Planned Behaviour by Taylor & Todd (1995)*

The following table maps out the DTPB constructs against the various context classifications found in the literature and relates them with the BYOD construct to assist in understanding the behaviour of mobile workers

| DTPB Construct | Context | BYOD Construct | Definition |
|---|---|---|---|
| Attitude (Towards IS use) – Perceived Advantage (Usefulness) | Personal Context | Mobility and BYOD enables productivity | Perceived usefulness refers to real user's subjective belief that using a specific system will increase job performance |
| Attitude – Simplicity of Use (ease of use) | Personal Context | BYOD is easier to use than Organization provided devices | Ease of use refers to the degree to which the real user believes the use of the target system to be free of effort |
| Attitude – Compatibility (values, experience and Needs) | Personal Context | BYOD provides mobile workers with productivity and satisfaction | Compatibility refers to the degree to which the user is engaged with a specific system according to their values, experience and needs |
| Subjective norm - Peer's Influence | Social Context | Every Mobile knowledge worker uses their BYOD to perform (knowledge) work | Peer's influence refers to social influence which plays a key role in user's adoption of a technology, in this case BYOD. |
| Subjective norm - Superior's Influence | Organizational /Business Context | Lack of Superior's influence, such as Lack of Policy enforcement (Organisation) contributes to behaviours leading to knowledge leakage risk. | Superior's influence refers to the effect that rules and authority has on user's intention to use technology. In this case, the absence of security guidelines on the use of BYOD in organizations (superior's influence) makes employees engage in risky behaviours. |
| Perceived Behavioural Control - Self-Efficacy | Technology Context | BYOD enables work efficacy | Self-efficacy refers to the perceived ability by a user that the use of a system is effective |
| Perceived Behavioural Control - Resource Facilitating condition | Technology Context/ Geographical context / Physical Context | BYOD are powerful devices which provides the resources and tools to perform knowledge work easily | Resource facilitating condition is a subjective factor and refers to the degree to which an individual believes that an organizational infrastructure exists to support the use of a system. |
| Perceived Behavioural Control - Technology Facilitating condition | Technology Context | Network infrastructure and mobile computing facilitate the use of BYOD | Technology facilitating condition refers to the degree to which an individual believes that a technical infrastructure exists to support the use of a system. |

*Table 3 DTPB Constructs vs. Context Types*

## 3   Conceptual Research Framework

Figure 2 depicts the proposed conceptual framework of this research extended from the DTPB theoretical model, which illustrates that knowledge leakage can occur due to the underlying factors determined by DTPB and it is also affected by the use of BYOD and usage contexts on mobile worker behaviour which directly elicits accidental knowledge leakage. The conceptual framework also shows how the perceived behavioural control construct, which exerts on both intentional and accidental user behaviour, directly impacts *accidental* knowledge leakage.

The BYOD construct in the framework refers to the mobile device use factor which increases knowledge leakage risk outside the boundary of the organization (boundary-spanning technology) e.g., the information security policy enforcement is more challenging to manage when the device is employee owned and it is beyond the company's network perimeter as opposed to a workstation provided by the organization. Hence, this determinant affects the way workers engage in activities that lead to knowledge leakage risks.

On the other hand, the context construct relates to the mobile device usage context, defined previously, and refers to the combination of environment, device with its content, and the mobile worker, this influences the way the employee interacts with their surroundings and the device, determining their behaviour, for example, the behaviour of a mobile worker within the organization is drastically different to a mobile worker in a train or an airport, thus the risk profiles change according to the context.

As shown in figure 2, the intention (outcome of DTPB), BYOD and context, all these three constructs, influence the mobile worker behaviour which, in turn, elicits accidental knowledge leakage when the employee inadvertently leaks organizational knowledge by disregarding security policies in order to get the work done, e.g., connecting to a public open wireless network from an airport to send a sensitive file containing organizational trade secrets or intellectual property to a work colleague.

At present, the extended constructs of the conceptual framework provide a foundation to understand the factors underlying the behaviour which results in accidental knowledge leakage rather than a fully –fledged mechanism that can measure behaviour directly. The framework aim to explain how various influence factors can have an impact upon a mobile worker and how these affect the final accidental behaviour. Even though, it is not possible to accurately pinpoint the exact issue behind the behaviour of a person, an organization could use the framework to understand and relate the factors that are likely to influence mobile employees. If the organisation is able to gauge these, then it may gain insight into the level of security-compliant behaviour to be expected, and the extent to which it may need to tune controllable factors such as technical preventive controls, information security policies (ISP), and security education training and awareness (SETA) programs to compensate.

The conceptual framework rather than a practical tool is an assessment instrument, whose role is to assist in understanding of just how many of the behavioural influences can realistically be considered to fall within the organisation's control. For example, it helps to illustrate the extent to which formal security education, training and awareness initiatives may have to compete against the influences exerted by various other factors. It also highlights that there are various factors that the organisation has no real opportunity to influence, but which might consequently need to be accounted for in the strength of other messaging.

Additionally, the conceptual framework will be subject to initial validation via multidisciplinary focus group, including a mixture of IT, security and human resource professionals and related academic staff. The group will be presented with an initial version of the framework to discuss around the validity of the factors included and whether any significant influences have been omitted or new constructs need to be added.

Likewise, this study adopts a qualitative approach, specifically, is classified as exploratory and explanatory since it seeks to explore the mobile behaviour among workers that may contribute to leakage of knowledge through the use of BYOD and explain the factor or influences that underlie this social IS behaviour. Therefore, it will use a combination of approaches such as: semi-structured interviews, carefully designed using the DTPB, that will be recorded to ensure accuracy of response; direct observations to analyse workers' behaviour; document reviews such as policies, procedures, guidelines, manuals, among others; and multiple case studies conducted in knowledge intensive organizations, such as consulting firms, software companies, pharmaceutical and service institutions, among others, to explore and explain strategies used by these firms to protect their confidential information and knowledge from being leaked through BYOD.

Upon completion of this research, the actual reasons of mobile workers leaking organizational knowledge through the use of BYOD and the strategies utilized by organizations to prevent it will materialize. The reasons are expected to revolve around the factors determined by the conceptual framework based upon DTPB model. Furthermore, a maturity model as an assessment tool to gauge knowledge leakage risk in organizations through BYOD will be developed to assist in the design of effective prevention methods such as SETA programs, information security policies and other forms of control mechanisms.

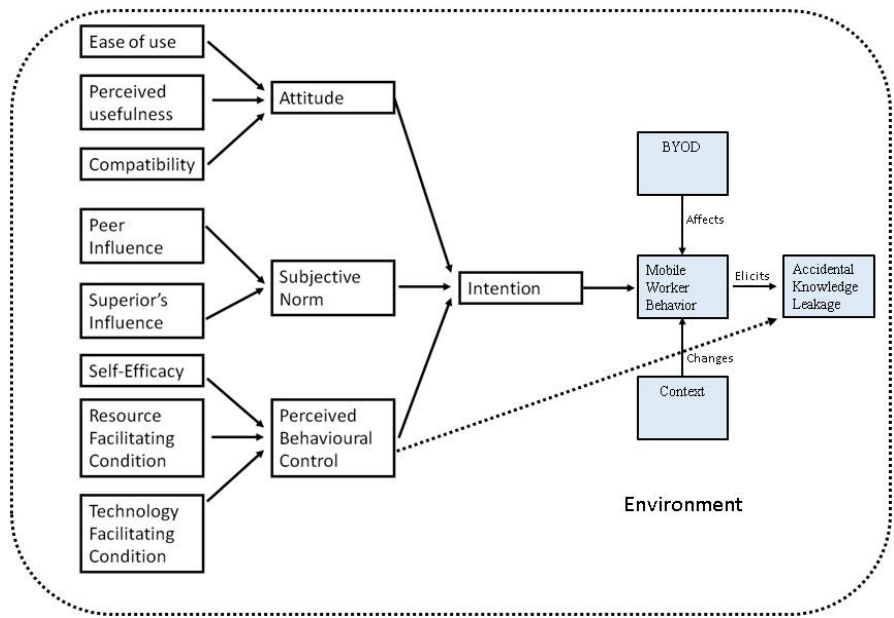

*Figure 2 Conceptual research Framework*

## 4　Conclusion and Future Work

Knowledge is a corporate asset and drives competitive advantage as it is the source of innovation in organizations, as such it is important that organizations understand the factors behind knowledge leakage in order to address and develop better policies and SETA (Security Education Training and Awareness) programs to prevent this risk from happening.

This study reviewed theories from different disciplines such as sociology, psychology and criminology to understand the reason behind BYOD knowledge leakage behaviour and found that DTPB is the most suitable theory as it exerts on its perceived behavioural control construct that behaviour can be intentional and accidental. This finding agrees with the insider threat literature in that accidental security incidents by insiders happen more often and could have greater potential for harm than malicious insider attacks (Colwill 2009). Therefore, as part of the future work for this research, multiple case studies will be conducted to validate and evaluate these constructs within knowledge intensive organizations that have adopted BYOD for their employees.

Thus, the key contribution of this study is to provide the reasons behind the behaviour of mobile knowledge workers that elicit knowledge leakage through BYOD and therefore assist organizations with a better design and implementation of information security policies addressing specifically BYOD and SETA programs to mitigate this issue. Likewise, the study will seek to offer contributions to the knowledge security and mobile security research and practice, addressing the gaps concerning behavioural facets.

Finally, as knowledge leakage is a complex topic and involves the human nature dimension, the need for further research in tacit knowledge management and organizational strategies to holistically address knowledge security should be addressed as it is still in its infancy (Manhart and Thalmann 2015). Furthermore, future studies on the adaptation and application of maturity models to assess knowledge leakage risk in organizations to facilitate the assessment of their current 'as-is' status and provide a strategic road map to a 'to-be' state and validation through multiple case studies are part of the future work of this research.

## 5　References